\documentclass[aps,pre,one,11pt]{revtex4}%
\usepackage{epsfig}
\usepackage{amsmath}
\usepackage{amsfonts}

\begin{document}

\title{Thermodynamics and dynamics of
the formation of spherical lipidic vesicles}

\author{E. \surname{Hern\'andez} Zapata}
\affiliation{Departamento de F\'isica, Matem\'aticas e
Ingenier\'ia, Universidad de Sonora, H. Caborca, C.P. 83600, Sonora, M\'exico.}
\author{L. \surname{Mart\'inez} Balbuena}
\affiliation{Departamento de Investigaci\'on en F\'isica, Universidad de
Sonora, Apartado Postal 5-088, 83190 Hermosillo, Sonora, M\'exico.}
\author{I. \surname{Santamar\'ia} Holek}
\affiliation{Facultad de Ciencias, Universidad Nacional Aut\'{o}noma de
M\'{e}xico. Circuito exterior de Ciudad Universitaria. 04510, D.
F., M\'{e}xico.}

\begin{abstract}
We propose a free energy expression accounting for the formation
of spherical vesicles from planar lipidic membranes and derive a
Fokker-Planck equation for the probability distribution describing
the dynamics of vesicle formation. We found that formation may
occur as an activated process for small membranes and as a
transport process for sufficiently large membranes. We give
explicit expressions for the transition rates and the
characteristic time of vesicle formation in terms of the relevant
physical parameters.
\end{abstract}

\maketitle
\section{Introduction}\label{sec:intro}
Phospholipid vesicles have been widely used as model systems for
studying the dynamics and structural features of many cellular
processes, such as endocytosis \cite{lei}, exocytosis \cite{cans},
cell fusion \cite{pantazatos, lei1}, transport and diffusion
phenomena \cite{castro}, and membrane elastic properties
\cite{svetina}. In addition to its importance for basic research
in the biological sciences, closed vesicles (liposomes) has been
used as vehicles for the encapsulation of macromolecules such as
nucleic acids \cite{leonetti, renneisen} as well as polymers and
small molecules \cite{dominak}. Large enough vesicles can be
individually manipulated with a glass micropipette \cite{evans,
ipsen}, and the vesicle membrane rigidity and, in general,
membrane elastic properties can be measured \cite{evans1,longo}.
They have also been used as microreactors useful in the study of
chemical reactions in geometrically confined spaces
\cite{karlsson, bolinger}. In general, lipidic vesicles constitute
nanocontainer systems ideally suited for the isolation,
preservation, control and transport of a small number of
molecules.

There is a variety of experimental methods to prepare phospholipid
vesicle suspensions, reviewed in \cite{lasic}. One of the most
widely used methods is the hydration of a dry phospholipid film
\cite{lopez, paredes}, resulting spontaneously in a population of
multilamellar vesicles with a high polydispersity in sizes and
shapes.  On the other hand, the formation of a unilamellar vesicle
usually involves an intermediate structure in the form of a planar
bilayer fragment, which is unstable, due to its exposed edges.
These small planar bilayers can be grown by detergent depletion,
phospholipid precipitation or they can be formed from pre-existing
bilayers \cite{lasic}.  It is possible to prepare a population of
giant unilamellar spherical vesicles when a dry phospholipid film
is hydrated in the presence of an AC electric field
\cite{angelova, dimitrov}. The resulting vesicle radius can be as
high as 50 micrometers. A similar effect is exhibited by charged
phospholipids. The bilayers ionize upon contact with water and
they swell due to the repulsion of the bilayers, leading to the
spontaneous formation of unilamellar vesicles \cite{tardieu}.

In spite of the large experimental work existing, to our knowledge
there is no systematic theoretical model describing the dynamics
of formation of a unilamellar spherical vesicle from a small
planar membrane. Such a model could be useful for the
characterization and control of the vesicle formation process and
it could be tested by performing single-vesicle simulations and
experiments. For instance, the video microscopy analysis of the
closing dynamics of laser-generated transient pores on
phospholipid membranes \cite{srividya} could be very useful in
this regard.

In this article, we propose a simple theoretical model for
spherical-vesicle formation from a planar membrane, assuming that
membrane rigidity and edge tension are the main contributions. We
first calculate the free energy cost of vesicle formation and
then, using this free energy and the rules of mesoscopic
nonequilibrium thermodynamics (MNET), we derive a Fokker-Planck equation
governing the evolution in time of a nonequilibrium distribution
function that depends on time and the mesoscopic variable
characterizing the instantaneous state of the system. Our analysis
leads to identify that the ratio between the contour energy to
curvature energy determines two main mechanisms of vesicle
formation: \emph{i)} An activated process for small values of the
energy ratio and \emph{ii)} a transport process for values larger
than a critical value of the energy ratio. A detailed analysis of
these two cases is performed  leading to explicit relations for
the vesicle formation rates in the first case and for the
characteristic formation time in the second one. Our analysis is
complemented with a numerical solution of the Fokker-Planck
equation.

MNET has been also used in other nanometric processes where curvature and
surface tension effects are the main driving forces, such as matter
agglomeration systems; see for example \cite{gadomski2003}.
The effect of linear tension on growth morphologies in 2D
has been also studied in \cite{gadomski2002}, where the entropy
production has been shown to be the dominant selection mechanism.

The article is organized as follows. Sec. \ref{sec:model} is devoted to derive
the expression for the free energy cost of vesicle formation by
using equilibrium arguments. In section \ref{sec:dinamic}, we use this free
energy to formulate a kinetic model for vesicle formation in terms
of a Fokker-Planck equation and to analyze its implications.
Finally, in section \ref{sec:conclusion} we present conclusions.

\section{The free energy}\label{sec:model}
In this section we formulate a simple model for the free energy
associa\-ted to a phospholipid membrane in the process of wrapping
in order to form a spherical vesicle.

\begin{figure}
    \centerline{\includegraphics[width=12pc]{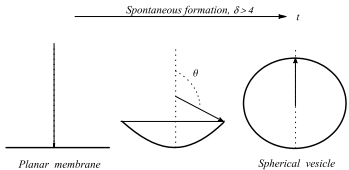}}
    \caption{Schematic representation of vesicle formation from a planar membrane.}
\label{velpromedio1}
\end{figure}

We will assume that in every stage of the process the membrane
adopts the form of a spherical bowl as shown in Figure
\ref{velpromedio1}. In this process, we will consider two
competing energies, one $F_B$ associated with the bending of the
membrane that favours planar membranes and another one $F_l$ due
to the contour of the membrane which favours spherical vesicles.

According to the well established Helfrich theory the free energy
of bending per unit area\textbf{, $f_B$,} obeying the relation
$F_B=\int f_B dA$, associated to a local
deformation of a membrane is given by \cite{helfrich,safran}
\begin{equation}\label{fB1}
f_B= 2 \kappa (H-c_0)^2 + \overline{\kappa} K,
\end{equation}
where $\kappa$ and $\overline{\kappa}$ are the bending and the
saddle-splay moduli respectively, $c_0$ is the spontaneous
curvature of the bilayer. Here $H = (1/2)(c_1+c_2)$ is the mean
curvature, $K =c_1 c_2$ is the Gaussian curvature and $c_1$ and
$c_2$ are the local principal curvatures of the system. Since we
are interested in homogeneous bilayers, then we may assume $c_0
=0$. In our bowl approximation, both principal curvatures are
identical and equal to inverse radius of the sphere $r$:
$c_1=c_2=1/r$. Therefore, the bending free energy simplifies to
\begin{equation}\label{fB2}
F_B= A \frac{\kappa_b}{r^2} ,
\end{equation}
where $\kappa_b=2 \kappa + \overline{\kappa}$ and $A$ is the area
of the membrane which will be assumed as constant. The contour
free energy has the simple form
\begin{equation}\label{fL1}
F_l= \gamma l,
\end{equation}
where $\gamma$ is the edge tension and $l$ is the contour length.
The total free energy is the sum of both contributions
$F=F_B+F_l$. Eqs. (\ref{fB2}) and (\ref{fL1}) can be rewritten in
terms of the angle $\theta$ (see Figure \ref{velpromedio1}),
leading to the following expression for the free energy
\begin{equation}\label{total}
F(\theta)= 2 \pi \kappa_b \left[ 1+ \cos(\theta)\right]+2 \gamma \,(\pi \,A)^{1/2} \sin(\theta/2).
\end{equation}
To derive this equation, we have used the fact that the total area
of the bowl is $A = 2 \pi r^2 \left[ 1+ \cos(\theta)\right]$, that
the contour length is given by $l=2 \pi r \sin(\theta)$, and used
the trigonometric relation $\sin(\theta)/\left[ 1+
\cos(\theta)\right]^{1/2}=\sqrt{2}\sin(\theta/2)$. For
convenience, we will use the following dimensionless form of the
free energy
\begin{equation}\label{totalAdim}
\tilde{F}(\theta)= \frac{F}{ 2 \pi \kappa_b}= 1+ \cos(\theta)+ \delta \sin(\theta/2),
\end{equation}
where $\delta =(A/\pi)^{1/2} (\gamma / \kappa_b)$.

\begin{figure}
    \centerline{\includegraphics[width=12pc]{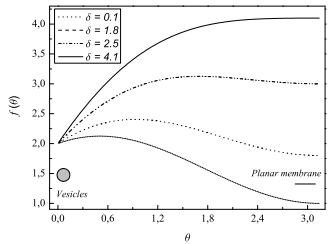}}
\caption{Dimensionless free
energy $\tilde{F}(\theta)$ as function of $\theta$ for $\delta=0.1, 1.8,
2.8, 4.1$.} \label{velpromedio2}
\end{figure}

In Eq. (\ref{totalAdim}) it is clear that the parameter $\delta$
determines the form of the free energy as a function of $\theta$,
and thus it determines when the planar membrane is stable and when
it will spontaneously form a closed vesicle. We can identify the
following three regimes (see Figure \ref{velpromedio2}):
\begin{enumerate}
    \item For $\delta\leq 2$, that is, when the linear tension is small
    compared to the bending constant $\kappa_b$, then the free energy
    has a minimum at $\theta =\pi$ that corresponds to a planar membrane.
    \item For $2<\delta<4$, there is a competition between contour
    and bending forces. As a result of this, the free energy has a minimum at $\theta =0$
corresponding to a closed spherical vesicle with an energy barrier
centered at $\theta^* =2 \arcsin(\delta/4)$. The free energy
difference with respect to the planar membrane is given by
\begin{equation}\label{totalAdim2}
\Delta \tilde{F}= \tilde{F}(\theta^*)-\tilde{F}(\pi)= \frac{1}{8}(\delta - 4)^2.
\end{equation}
     \item For $\delta>4$ there is no energy barrier and the closed
     spherical vesicles are formed spontaneously.
\end{enumerate}

Let us now estimate the possible values of $\delta$ for real
systems. For lipid bilayers the typical experimental values of the bending
modulus are $\kappa \sim 5 - 25 k_BT$ whereas for block copolymer
bilayers a typical value is $\kappa \sim 40 k_BT$, \cite{boal,Forster}.
The bilayer saddle-splay modulus is approximately given by
$\overline{\kappa}\sim - a \kappa$ with $a \sim 1$ or less
\cite{morti,kozlov}. Therefore, $\kappa_b \sim 5 - 25 k_BT$. The
edge tension is of the order of $\gamma \sim 1-2\,k_BT/nm$,
\cite{boal,Forster}.

For definiteness let us consider $\kappa_b \sim 25 k_BT$ and
$\gamma \sim 1k_BT/nm$ leading to minimum radius of the vesicles
(corresponding to $\delta =2$) of $r_{min}\sim \kappa_b/\gamma =
25nm$. For radius in the range between $r \sim 25 - 50nm$
an energy barrier has to be overcome in order to form vesicles
while for radius larger than \textbf{$50nm$} the vesicles will
form spontaneously.

\section{Dynamics of vesicle formation}\label{sec:dinamic}

At isothermal conditions, the free energy given in Eq.
(\ref{totalAdim}) can be interpreted (up to a constant) as the
energetic cost or the minimum work necessary to form a vesicle
\begin{equation}\label{Rmin}
R_{min}=2 \pi \kappa_b \left[ 1+ \cos(\theta)+ \delta \sin(\theta/2)\right].
\end{equation}

This quantity can be used to derive a Fokker-Planck equation for the
distribution function $P(\theta,t)$ of finding the membrane in a stage
characterized by $\theta$ at time $t$. This distribution function
is normalized and then satisfies a continuity equation of the form
\begin{equation}\label{continuity}
\frac{\partial P}{\partial t} = - \frac{\partial }{\partial \theta} (P v_{\theta}),
\end{equation}
where $P v_{\theta}$ is a diffusion probability current in
$\theta$-space.

The Fokker-Planck equation can be obtained by using the rules of
nonequilibrium thermodynamics \cite{degroot,reviewMNET, landau}.  This
objective can be achieved by first calculating the entropy
production of the system using the Gibbs entropy postulate
\cite{reviewMNET}
\begin{equation}\label{gibbs}
\Delta s = - k_B \int P \ln \frac{P}{P_{eq}} d\theta.
\end{equation}
In this equation $\Delta s$ is the entropy change in the process
of formation of the vesicle and the integration is carried out
over all the range of values of $\theta$ (from zero to $\pi$).
Here, $P_{eq}$ is the equilibrium reference distribution given by
\cite{landau1}
\begin{equation}\label{peq}
P_{eq}=P_0e^{-{R_{min}}/{k_BT}}=P_0e^{-2 \pi \kappa_b \left[ 1+ \cos(\theta)+ \delta \sin(\theta/2)\right]/{k_BT}},
\end{equation}
where $P_0$ is a normalization factor. When $\delta<2$, Eq.
(\ref{peq}) is a very narrow function centered around
$\theta=\pi$, thus implying that the system remains as a planar
membrane. For values of $\delta$ slightly larger than $2$, one
finds that the equilibrium state has a coexistence of vesicles and
planar membranes. Otherwise, closed vesicles are the preferred
configuration of the system.

Now, by taking the derivative of Eq. (\ref{gibbs}) with respect to
time and using Eq. (\ref{continuity}), we obtain for the time
derivative of the entropy:
\begin{equation}\label{sigma}
\frac{\partial s}{\partial t} = k_B \int_0^{\pi} \frac{\partial }{\partial \theta} (P v_{\theta}  \ln \frac{P}{P_{eq}})  d\theta
- \frac{1}{T}\int_0^{\pi} P v_{\theta} \frac{\partial \mu}{\partial \theta}
 d\theta.
\end{equation}
where we have defined the nonequilibrium chemical potential
$\mu=k_BT \ln|{P}/{P_{eq}}|$. This equation contains two terms,
the first one constitutes the entropy flow and the second one is the
entropy production $\sigma$, given by
$\sigma=({1}/{T})\int_0^{\pi} P v_{\theta} ({\partial \mu}/{\partial \theta}) d\theta$.
 From Eq. (\ref{sigma}) we may formulate
linear relationships between the current $P v_{\theta}$ and its
conjugated force $\frac{\partial \mu}{\partial \theta}$ in
the form
\begin{equation}\label{linearlaws}
P v_{\theta}= - \alpha P \frac{\partial \mu}{\partial \theta},
\end{equation}
where $\alpha$ is the corresponding Onsager coefficient satisfying
Onsager reciprocity relations \cite{degroot}. This use of a linear relationship
assumes that the process is not too far from equilibrium, and therefore it may
be not valid in general (this may be the case of charged vesicles). Note that for values
$\delta >2$, in the boundary $\theta = 0$, the velocity
$v_{\theta}$ should vanish since the free energy has a local
minimum. This condition does not affect the election
(\ref{linearlaws}) for other values of $\theta$.

Now, by substituting Eq. (\ref{linearlaws}) into
(\ref{continuity}), and using (\ref{peq}) we finally obtain
\begin{equation}\label{FP}
\frac{\partial P}{\partial t} =
\alpha \frac{\partial }{\partial \theta}
\left\{ 2 \pi \kappa_b  \left[-\sin(\theta)+\frac{\delta}{2}\cos\left(\frac{\theta}{2}\right)\right]P+
k_BT \frac{\partial P}{\partial \theta}\right\}.
\end{equation}
This is the Fokker-Planck equation governing the time evolution of
the probability distribution during the formation of the vesicle.
It contains a driving term characterized by the force $\partial F
/\partial \theta$ and a diffusion term characterized by the
diffusion coefficient $D=k_BT \alpha$. Here $\alpha$ plays the
role similar to that of a friction or mobility coefficient in
usual Brownian motion. In this case, it can be interpreted as a
parameter characterizing the viscous or friction forces exerted on
the membrane by the solvent or even it may include interlayer
friction \cite{Miao}. To make sure that $P(\theta,t)$ will remain
confined in the range $0<\theta\leq \pi$ the initial condition can
be written in the form
$P(\theta,0)=[H(\theta)-H(\pi-\theta)]P_0(\theta)$ with
$H(\theta)$ the Heaviside function.

\subsection{Vesicle formation in the presence of energy barriers}

The Fokker-Planck equation (\ref{FP}) can be used to calculate the
transition rates $r_p$ from planar membranes to spherical vesicles
(and for the inverse process $r_v$) in the regime where the energy
barrier controls the dynamics, that is, for values of $\delta$ in
the interval $\delta \in (0,4)$. This objective can be achieved by
following the usual methods of activated processes \cite{risken}.

Let us estimate the transition rate $r_p$ in the stationary state
assuming that the number of planar membranes is much larger than
the number of vesicles. This can be done by assuming a two-state
dynamics in the presence of a barrier. In this case the net
current
\begin{equation}\label{FP-stat}
j =  -\alpha \left\{ 2 \pi \kappa_b
\left[-\sin(\theta)+\frac{\delta}{2}\cos\left(\frac{\theta}{2}\right)\right]P+
k_BT \frac{\partial P}{\partial \theta}\right\} ,
\end{equation}
obtained from the Fokker-Planck equation (\ref{FP}), is a constant
in $\theta$-space, in the stationary case. Then, the transition
rate $r_p$ is defined by $r_p = j/ P_p$, where $P_p$ is the total
number of planar membranes that can be calculated by integrating
the stationary distribution function $P_s(\theta)$ from $\theta^*$
to $\theta=\pi$ \cite{risken}.

The explicit expression for $j$ and $P_p$ can be obtained by
expanding in a Taylor series up to second order in $\theta$ the
free energy potential $F$ about its maximum at $\theta^*$ and its
local minimum at $\theta=\pi$. This procedure yields the
approximate expressions
\begin{eqnarray}\label{totalAdim-Aprox}
F_{max}(\theta) \simeq \frac{\pi \kappa_b}{4}
\left(16+\delta^2\right) - \frac{\pi
\kappa_b}{16}(16-\delta^2)(\theta-\theta^*)^2,
\\
F_{\pi}(\theta) \simeq 2 \pi \kappa_b \delta + \frac{\pi
\kappa_b}{4}(4-\delta)(\theta-\pi)^2 \label{totalAdim-Aprox2}
\end{eqnarray}

\begin{figure}
    \centerline{\includegraphics[width=12pc]{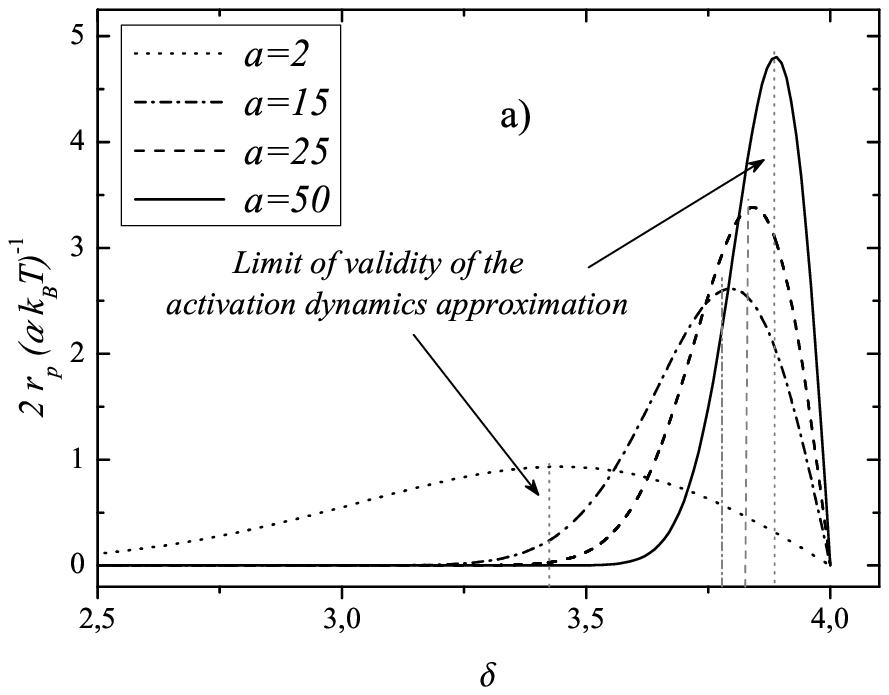}}
    \centerline{\includegraphics[width=12pc]{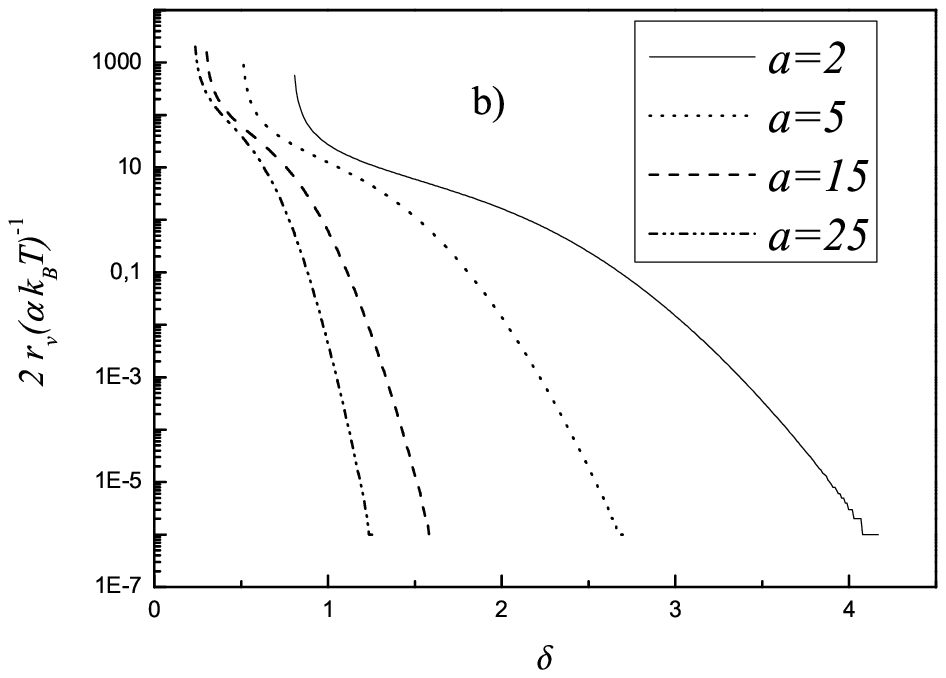}}
\caption{ a) Dimensionless
transition rate $r_p$ as a function of $\delta$ for different
values of $a=\kappa_b/k_BT = 2, 15, 25, 50$ at room temperature.
The vertical lines indicate the limit of validity of the
activation dynamics approximation. b) Dimensionless transition
rate $r_v$ as a function of $\delta$ for different values of $a=
2, 5, 15, 25$ at room temperature.} \label{rates}
\end{figure}

To calculate $j$ and $P_p$ one then uses Eqs.
(\ref{totalAdim-Aprox}) and (\ref{totalAdim-Aprox2}),
respectively. Now, the transition rate $r_p$ from planar membranes
to spherical vesicles is
\begin{eqnarray}\label{rate}
r_p = \frac{\alpha \kappa_b}{4} (4-\delta) (4+\delta)^{1/2}e^{-\pi
\kappa_b (\delta-4)^2/ 4k_BT}.
\end{eqnarray}
This expression is valid as long as the energy barrier is larger
than the thermal energy. According to our expression this imposes
the condition $\delta < 4 - \sqrt{\frac{4k_BT}{\pi \kappa_b}}$.
When this condition is not satisfied, the vesicle formation must
be analyzed as a transport process. In Figure \ref{rates}a we show
$r_p$ as a function of $\delta$ for different values of
${\kappa_b}/{k_BT}$.

In order to calculate the rate $r_v$ of the inverse process when
the initial condition is such that the number of vesicles is much
larger than the number of planar membranes, we first approximate
the free energy around the local minimum at $\theta=0$, obtaining
\begin{equation}
F_{0}(\theta) \simeq \frac{\pi \kappa_b}{4} \left(16+
\delta^2\right)- \pi \kappa_b \left(\theta -
\frac{\delta}{2}\right)^2.\label{totalAdim-Aprox3}
\end{equation}

Note that the quadratic term in the approximation of free energy
at $\theta=0$ is negative. Thus, when evaluating the number of
vesicles $P_v$ around this minimum we obtain
\begin{eqnarray}\label{nv}
P_v= \sigma_v
e^{-\frac{\pi \kappa_b}{4k_BT}\delta^2}\int^{\delta/2}_0{e^{\frac{\pi \kappa_b}{k_BT}
\left(\theta - \delta/2\right)^2}},
\end{eqnarray}
where $\sigma_v d\theta$ is the number of membranes between $0$
and $d\theta$. In this case, the integration of the Boltzmann
factor, $\exp(-F_{0}(\theta)/k_BT)$, must be evaluated between the
minimum and the position of the maximum at $\theta^*$. For
simplicity sake, we have approximated $\theta^*$ up to first order
in $\delta$; that is, $\theta^* \sim \delta/2$. After calculating
the integral our estimation for the transition rate $r_v$ from
spherical vesicles to planar membranes is
\begin{eqnarray}\label{rate2}
r_v = \frac{\alpha \kappa_b}{2}
\frac{(16-\delta^2)^{1/2}}{erfi\left(\frac{\delta}{2}\sqrt{\frac{\pi
\kappa_b}{k_BT}}\right)}.
\end{eqnarray}

In Figure \ref{rates}, we show the behavior of $r_{v}$ as a
function of $\delta$ for different values of $a= 2\pi \kappa_b
/k_BT$. For a given value of $a$, the transition rate grows as the
energy barrier decreases ($\delta$ increases). For a given value
of $\delta$, the transition rate depends on the relative value of
the bending energy with respect to the thermal energy, $a$. For
constant temperature $T$, decreasing the bending modulus
$\kappa_b$ favours the formation of vesicles associated with
increasing values of \textbf{$r_{v}$}. It is interesting to notice
that the Arrhenius law is a consequence of two ingredients: i) the
assumption of the linear law Eq. (\ref{linearlaws}) which implies
a process not too far from equilibrium, and ii) the fact that the
minimum of the potential is an extremum, see Eq. (\ref{rate}).
This second condition is not fulfilled in the case of Eq.
(\ref{rate2}) which clearly is not an Arrhenius type law.
Non-Arrhenius behaviors can also emerge as a consequence of other
mechanisms, see for example \cite{agustin1,agustin2}.

\begin{figure}
    \centerline{\includegraphics[width=12pc]{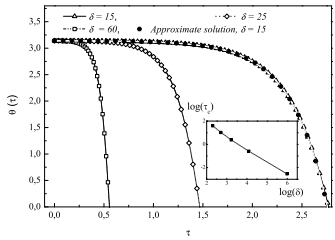}}
\caption{Angle $\theta$ as a
function of the dimensionless time $\tau=2\pi\kappa_b\alpha t$ for
the following values of $\delta$: 15, 25 and 60 obtained by
solving numerically Eq. (\ref{ecTheta}). The solid circles
represent an analytical solution given in the text for $\delta=
15$. The inset shows a log-log plot of vesicle formation time
$\tau_c$ versus $\delta$ for $\delta =$ 10, 15, 25, 60, 400. The
slope of the straight line is $\sim -1.07$.} \label{fig4}
\end{figure}

\subsection{Vesicle formation as a transport process}

For values of $\delta$ larger than 4, the absolute minimum of the
free energy occurs at $\theta = 0$ (corresponding to spherical
vesicles) without the pre\-sence of energy barriers. Therefore, in
this case the dynamics must be analyzed as a transport process.

To do this, one may neglect thermal fluctuations so that the
distribution function can be approximated by a Dirac delta
function $P(\theta,t)=\delta[\theta-\theta(t)]$, \cite{zwanzig}.
In this case, after multiplying by $\theta$ and integrating  over
all $\theta$-space, Eq. (\ref{FP}) reduces to the dynamical
equation
\begin{eqnarray}\label{ecTheta}
\frac{d }{d\tau}\theta(\tau) =- \frac{\delta}{2}
\cos\left(\frac{\theta}{2}\right)+ \sin\left({\theta}\right),
\end{eqnarray}
where we have defined the dimensionless time
$\tau=2\pi\alpha\kappa_b t$. Since this equation cannot be solved
analytically, we have solved it numerically by using a Runge-Kutta
method. The solutions (open symbols with lines) as a function of
$\tau$ for three different values of $\delta$ are shown in Fig.
\ref{fig4}. As initial condition we used $\theta(0)=3.14$
since this value represents a nearly planar membrane but with a
small perturbation that permits the membrane to evolve to its
equilibrium state (closed vesicle).

As it is clear from the figure, during most of the evolution time
the value of $\theta$ is close to $\pi$. Therefore, one may
approximate Eq. (\ref{ecTheta}) around $\theta=\pi$ to first order
in its series expansion. This can be solved with the same initial
condition leading to
$\theta(\tau)\sim\pi-0.0016\,\exp[(\delta-4)\tau/4]$. This solution
is represented in Fig. \ref{fig4} by the solid circles.
As can be seen, it constitutes an excellent approximation at all
times.

For large values of $\delta$, this approximation allows us to
estimate the characteristic vesicle formation time $t_c$, given by
$t_c ={2}/{\left( \pi \kappa_b \,\alpha \, \delta\right) }={2}/{\left(\sqrt{\pi
A}\,\alpha\, \gamma\right)}$. The inset of Fig. \ref{fig4} shows
that the dependence of $\log(\tau_c)$ as a function of
$\log(\delta)$ is linear with a slope close to $-1$.

We now note that the Onsager coefficient $\alpha$ is a mobility
that depends on the dynamic viscosity of the solvent $\eta$ and
the membrane area $A$. Since $\alpha^{-1}$ has dimensions of
energy multiplied by time, it must depend on the combination
$\alpha^{-1} \sim \eta A^{3/2}$, yielding
\begin{eqnarray}\label{t-c}
t_c \sim\frac{2}{\sqrt{\pi}}\frac{\eta A}{\gamma}.
\end{eqnarray}
This relation predicts that the formation time is given by the
ratio between the friction force $\sim \eta A$ exerted by the
solvent on the membrane by the linear tension force $\gamma$ at
the contour of the membrane.

\begin{figure}
    \centerline{\includegraphics[width=12pc]{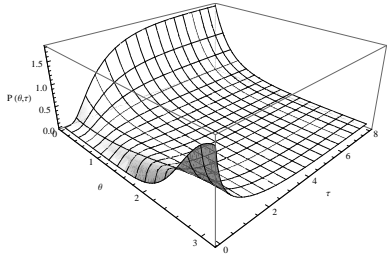}}
\caption{Probability
distribution $P$ as a function of angle and dimensionless time
$\tau$ obtained by numerically solving Eq. (\ref{FP}) with an
initial condition given by a Gaussian distribution centered at
$\theta=\pi$. At short times ($\tau<1$) diffusion dominates
spreading the distribution. For times $\tau>1$ the drifting force
dominates and the distribution becomes narrow close to
$\theta=0$.} \label{fig5}
\end{figure}

For typical values of vesicle area $A\sim 10^5 nm^2$, the dynamic
viscosity of the solvent $\eta \sim 10^{-3} Pa\,s$ and the linear
tension $\gamma \sim 1-2 k_BT/nm$, one obtains that the
characteristic formation time is of the order $t_c \simeq 1\,ms$.

Finally, we have numerically solved Eq. (\ref{FP}) in order to
study the effects of thermal fluctuations. We have taken
$\delta=5$ and $k_BT/2\pi\kappa_b = 1$. As an initial condition we
have taken a Gaussian distribution centered at $\theta=\pi$. At
short times ($\tau<1$) diffusion dominates spreading the
distribution whereas for times $\tau>1$ the drifting force
dominates and the distribution becomes narrow close to $\theta=0$.
These results are shown in Fig. \ref{fig5}.

\section{Conclusions}\label{sec:conclusion}

In this article, we proposed a free energy expression accounting
for the formation of spherical vesicles from planar membranes.
This energy depends on a single state variable and contains two
physical parameters related to the membrane rigidity and to the
edge tension. The equili\-brium properties of this energy depend on
the ratio, $\delta=(\pi A)^{1/2} \gamma /\kappa_b$, between the
contour energy and the Helfrich curvature energy. When $\delta <
4$ the free energy presents a barrier which disappears for larger
values of $\delta$.

Using mesoscopic nonequilibrium thermodynamics rules and the
equilibrium information, we have derived a Fokker-Planck equation
for the probability distribution describing the dynamics of
vesicle formation.

Two cases have been analyzed: \emph{i)} Formation in the presence
of barriers ($\delta < 4$) and \emph{ii)} formation as a transport
process ($\delta > 4$). In the first case we have derived
expressions for the transition rates of formation of vesicles from
planar membranes ($r_p$) and viceversa ($r_v$). Our expression for
$r_p$ follows an Arrhenius law [see, Eq. (\ref{rate})] and is an
increasing function of $\delta$. The rate of the inverse process
$r_v$ has an unusual dependence on temperature [see, Eq.
(\ref{rate2})] due to the fact that the free energy minimum at
$\theta=0$ (vesicles) is not an extremum. We have found that
$r_v/r_p$ is orders of magnitude smaller than one thus implying
that the unwrapping of the spherical vesicles is a very improbable
process even in the case when the free energy favours it.

In the second case, the free energy minimum always corresponds to
spherical vesicles and can be analyzed by using a deterministic
equation for the angle as a function of time after neglecting the
effects of thermal fluctuations. A simple analytical expression
that is an excellent approximation of the numerical solution
allows us to estimate the characteristic vesicle formation time
$t_c$ which is proportional to the membrane area and the viscosity
of the solvent, and inversely proportional to the edge tension
$\gamma$, see Eq. (\ref{t-c}). For typical values of phospholipid
membranes of linear dimensions $\sim 100nm$, we have obtained that
$t_c \sim 1ms$.

These results suggest that in typical experiments involving
video-based measurements, the dynamics of vesicles formation is
dominated by a transport process whereas for numerical
simulations, in which the studied system is small, the presence of
energy barriers could be relevant to the dynamics.

It could be interesting to test the present model by performing
single vesicle experiments and simulations in which the detailed
evolution in time of the membrane edge can be followed, so that
the characteristic vesicle formation time can be obtained.

The proposed model could be useful in the understanding of the
mechanisms of phospholipid vesicle formation widely used as model
experimental systems to study the thermoelastic properties of
cellular membranes.

\begin{acknowledgements}
We acknowledge useful discussions with Dr. A. Maldonado, and with
G. Paredes and C. Luna. This work has been done in the frame of
the Programa de Intercambio Acad\'emico UNAM-UNISON. We also thank
financial support by Grant No. DGAPA-IN102609.
\end{acknowledgements}

\end{document}